\documentclass[conference]{IEEEtran}
\IEEEoverridecommandlockouts
\usepackage{cite}
\usepackage{amsmath,amssymb,amsfonts}
\usepackage{algorithmic}
\usepackage{graphicx}
\usepackage{textcomp}
\usepackage{xcolor}
\usepackage{subfigure}
\usepackage[colorlinks,
            linkcolor=blue,
            anchorcolor=blue,
            citecolor=blue]{hyperref}
\usepackage{indentfirst}
\usepackage{mathrsfs}
\newtheorem{remark}{Remark}
\usepackage{multirow}
\usepackage[ruled]{algorithm2e}
\usepackage{multirow} 
\usepackage{bm}
\usepackage{parskip}
\usepackage{url}
\def\BibTeX{{\rm B\kern-.05em{\sc i\kerwn-.025em b}\kern-.08em
    T\kern-.1667em\lower.7ex\hbox{E}\kern-.125emX}}

\title{Max-Min Fairness and PHY-Layer Design of Uplink MIMO Rate-Splitting Multiple Access with Finite Blocklength}

\begin{document}
\author{

\IEEEauthorblockN{Jiawei~Xu, ~Bruno~Clerckx, \IEEEmembership{Fellow, IEEE}\vspace{-3.2em}}
\thanks{J. Xu is with Imperial College London. B. Clerckx is with the Department of Electrical and Electronic Engineering at Imperial College London, London SW7 2AZ, UK. (email: {j.xu20,b.clerckx}@imperial.ac.uk.}
}
\maketitle

\begin{abstract}
Rate-Splitting Multiple Access (RSMA) has emerged as a potent and reliable multiple access and interference management technique in wireless communications. While downlink Multiple-Input Multiple-Ouput (MIMO) RSMA has been widely investigated, uplink MIMO RSMA has not been fully explored. In this paper, we investigate the performance of uplink RSMA in short-packet communications with perfect Channel State Information at Transmitter (CSIT) and Channel State Information at Receiver (CSIR). We propose an uplink MIMO RSMA framework and optimize both precoders and combiners with Max-Min Fairness (MMF) metric and Finite Blocklength (FBL) constraints. Due to the high coupling between precoders and combiners, we apply the Alternating Optimization (AO) to decompose the optimization problem into two subproblems. To tackle these subproblems, we propose a Successive Convex Approximation (SCA)-based approach. Additionally, we introduce a low-complexity scheme to design the decoding order at the receiver. Subsequently, the Physical (PHY)-layer of the uplink MIMO RSMA architecture is designed and evaluated using multi-user Link-Level Simulations (LLS), accounting for finite constellation modulation, finite length polar codes, message splitting, adaptive modulation and coding, and Successive Interference Cancellation (SIC) at the receiver. Numerical results demonstrate that applying RSMA in uplink MIMO with FBL constraints not only achieves MMF gains over conventional transmission schemes such as Space Division Multiple Access (SDMA) and Non-orthogonal Multiple Access (NOMA) but also exhibits robustness to network loads. The benefits of splitting messages from multiple users are also illustrated. LLS results confirm the improved max-min throughput benefits of RSMA over SDMA and NOMA.

\end{abstract}

\section{Introduction}
It is agreed that the fifth-generation (5G) communication technology needs to be capable of supporting three generic services: massive Machine-Type Communications (mMTC), Ultra-Reliable and Low-Latency Communications (URLLC), and enhanced Mobile Broadband (eMBB). Among these three services, URLLC is required to provide the service with the highest reliability, i.e. $99.999\%$, and the lowest latency, i.e. less than $1$ ms \cite{huawei, 38913}. In this sense, systems must function with short-packet due to the low-latency requirements of URLLC, which makes it necessary to analyze system performance using Finite Blocklength (FBL) codes. URLLC demands FBL techniques to reduce latency while ensuring high reliability. Specifically, the authors of the groundbreaking study \cite{polyanskiy2010channel} have paved the way for theoretically analyzing the performance with FBL codes by laying out the fundamental bounds for the achievable rate with a given blocklength and error probability in Additive White Gaussian Noise (AWGN) channels. The achievable rate expression for FBL communications is then extended to Single Input Single Output (SISO) channels in \cite{polyanskiy2011scalar} and Multiple Input Multiple Output (MIMO) channels in \cite{yang2014quasi}. In addition, when multiple URLLC users connect to the same base station (BS), resource allocation becomes crucial due to resource constraints. Therefore, studying Multiple Access (MA) techniques with FBL constraints becomes imperative in this context. 

Among state-of-the-art downlink MA techniques, SDMA is a practical approach supported by the MIMO technology \cite{lu2014overview, goldsmith2003capacity}. It has been comprehensively approved that SDMA, to a significant extent, simplifies the signal processing complexity while enhancing sum-rate by scheduling users with orthogonal channels \cite{weingarten2006capacity,pan2004generalized, christensen2008weighted}. 
SDMA relies on the availability of accurate Channel State Information at Transmitter (CSIT). Therefore, imperfect CSIT in practical wireless communication networks becomes a primary bottleneck to achieve higher data rates in SDMA. Besides, SDMA is also sensitive to network load and it is only suitable in underloaded systems \cite{mao2022rate}. Apart from SDMA, Non-orthogonal Multiple Access (NOMA) has been recognized as another technique to increase the system capacity and reduce latency in wireless communication networks \cite{saito2013non,dai2015non}. A general framework of MIMO NOMA is to group users into different groups and apply Successive Interference Cancellation (SIC) to decode the intra-group interference while inter-group interference is treated as noise \cite{mao2018rate}. Nevertheless, decoding the message of other users at the intended user would deteriorate the Degree of Freedom (DoF) and would be inefficient in MIMO settings \cite{mao2018rate, clerckx2021noma}. 

Rate-Splitting Multiple Access (RSMA) has recently become a viable and effective MA and interference management technique to overcome the limitations of SDMA and NOMA \cite{mao2018rate, clerckx2023primer}. In downlink channels, this is done by splitting the user messages into common and private parts at the transmitter. The common parts are combined and encoded into one common stream and the private parts from each user are encoded into multiple private streams. As such, RSMA bridges SDMA and NOMA by partially decoding the interference and partially treating interference as noise, while SDMA fully treats interference as noises and NOMA fully decodes interference. Numerous works have demonstrated that RSMA improves the spectral and energy efficiency under both perfect and imperfect CSIT \cite{mao2018rate,clerckx2023primer,mao2019rate, mao2022rate,joudeh2016robust}. Moreover, Max-Min Fairness (MMF), as another vital metric, has been improved by RSMA \cite{joudeh2017rate} in multi-user downlink communications. Importantly, RSMA has also been shown to be a general and unified MA scheme that subsumes SDMA and NOMA as special instances of RSMA\cite{clerckx2019rate}.

Apart from its efficiency in the downlink, recent studies have proved that the system performance for uplink scenarios can be enhanced by adopting RSMA at the user side. The capacity region of a $K$-user Gaussian Multiple Access Channel (MAC) can be achieved by uplink RSMA, considering it as $2K-1$ virtual channels through message splitting \cite{rimoldi1996rate}, while NOMA requires joint decoding \cite{schiessl2020noma} or time-sharing which demands stringent synchronization due to the need to coordinate the transmissions of all users \cite{rimoldi1996rate} to achieve the capacity region. It has been demonstrated that uplink RSMA can not only improve the capacity in SISO channel \cite{yang2020sum} but also the MMF in SIMO channel \cite{zeng2019ensuring}. Moreover, RSMA with network slicing has been stated to provide a more flexible decoding order and a larger achievable rate region than Orthogonal Multiple Access (OMA) and NOMA \cite{liu2023network}. Uplink MIMO RSMA with electromagnetic exposure constraints has been studied in\cite{jiang2023rate}, demonstrating the significant gain over NOMA and SDMA in terms of sum-rate.

Although the aforementioned studies are based on the ideal assumption of Infinite Blocklength (IFBL), the investigation of RSMA with FBL constraints for both downlink and uplink has been done recently. For downlink communication with FBL, the performance of sum-rate and MMF has been notably improved by RSMA compared to SDMA in both underloaded and overloaded cases \cite{xu2022rate,xu2022max}. For uplink transmission with FBL, \cite{xu2022rate2} has shown that RSMA in the MU-SISO scenario can greatly reduce error probability with the same transmission rates as NOMA to improve the reliability of the transmission. RSMA has also been proven to obtain a large MMF gain over NOMA and Treat-Interference-as-Noise (TIN) systems and reduce latency significantly \cite{xu2024max}. However, \cite{xu2022rate2, xu2024max} primarily focused on uplink SISO channels without considering the benefits brought by MIMO. Therefore, it remains an important open problem to investigate how much gain RSMA can provide over its competitors in uplink MIMO systems with FBL. 

\subsection{Motivations and Contributions}
While downlink RSMA has been shown to provide significant gains in MIMO settings and FBL over SDMA and NOMA \cite{mishra2021rate,xu2022max,xu2022rate}, there has not been any work on uplink RSMA for MIMO and FBL. Work \cite{xu2022rate2} paved the way towards that goal but is limited to SISO. Therefore, our paper fills this gap thoroughly by providing theoretical foundations and verifications through link-level simulations. The contributions of the paper are summarized as follows:
\begin{itemize}
\item This paper establishes a general $K$-user uplink MIMO RSMA system in the finite blocklength regime. This framework is general in the sense that any arbitrary user can have multiple streams with each being split into two parts and transmitted to the BS. This is the first work investigating the performance of uplink MIMO RSMA systems with FBL constraints.
\item Based on the established system model, we formulate a MMF optimization problem with FBL constraints by designing the precoders of the split messages for each user and combiners at the BS. Because of the high coupling between precoders and combiners, we utilize Alternating Optimization (AO) by decomposing the original problem into two subproblems which are precoder optimization and combiner optimization, respectively. We propose a Successive Convex Approximation (SCA)-based method to address the formulated subproblems. This is the first work that optimizes both precoders and combiners for uplink MIMO RSMA systems with FBL constraints.
\item With the assumption of perfect CSIT and CSIR, numerical results demonstrate that in the $K$-user uplink MIMO, the fairness of FBL RSMA with optimized precoders and combiners outperforms NOMA and SDMA schemes even when splitting the messages of only one user. This contrasts with the convention of splitting the messages of $K-1$ users as outlined in \cite{rimoldi1996rate}. Furthermore, by increasing the number of splitting users, RSMA with the FBL constraints can outperform NOMA with IFBL settings. 
\item We illustrate the performance of RSMA in practical systems by developing the Physical (PHY)-layer architecture of RSMA with finite constellation modulation techniques, finite length polar codes, and Adaptive Modulation and Coding (AMC). Through such Link Level Simulations (LLS), we demonstrate that RSMA outperforms SDMA and NOMA in uplink MU-MIMO systems in terms of max-min throughput. This is the first study to develop the PHY-layer architecture and assess uplink RSMA performance using LLS.
\end{itemize}

\subsection{Organisation}
The rest of the paper is organized as follows. The system model of uplink RSMA in MU-MIMO system with FBL is specified in Section \ref{2}. The problem is formulated and optimized in Section \ref{3} including the convergence analysis of the proposed algorithm. Section \ref{4} discusses the PHY-layer architecture for MIMO RSMA. Simulation results are presented in Section \ref{5} and we conclude the paper in Section \ref{6}. 

\subsection{Notations}
Italic, bold lower-case, bold upper-case and calligraphic letters denote scalars, vectors, matrices and sets respectively. $(\cdot)^{H}, \text{tr}(\cdot), \lVert\cdot\rVert$ denote conjugate transpose, trace and Euclidean norm of the argument. $\mathbb{C}^{M\times N}$ denotes the set of all $M\times N$ dimensional matrices with complex-values entries. $|\cdot|$ denotes the absolute value if the argument is a scalar or the cardinality if the argument is a set. $\cdot\setminus\cdot$ denotes the difference between two sets if the arguments are sets. $\mathcal{CN}(\mu, \sigma^{2})$ denotes circular symmetric complex Gaussian distribution with mean $\mu$ and variance $\sigma^{2}$. $\mathbb{F}_{2}^{{K}}$ represents the binary field of dimension $K$.

\section{System Model}\label{2}
\begin{figure*}[t!]
    \centering
    \includegraphics[scale=0.45]{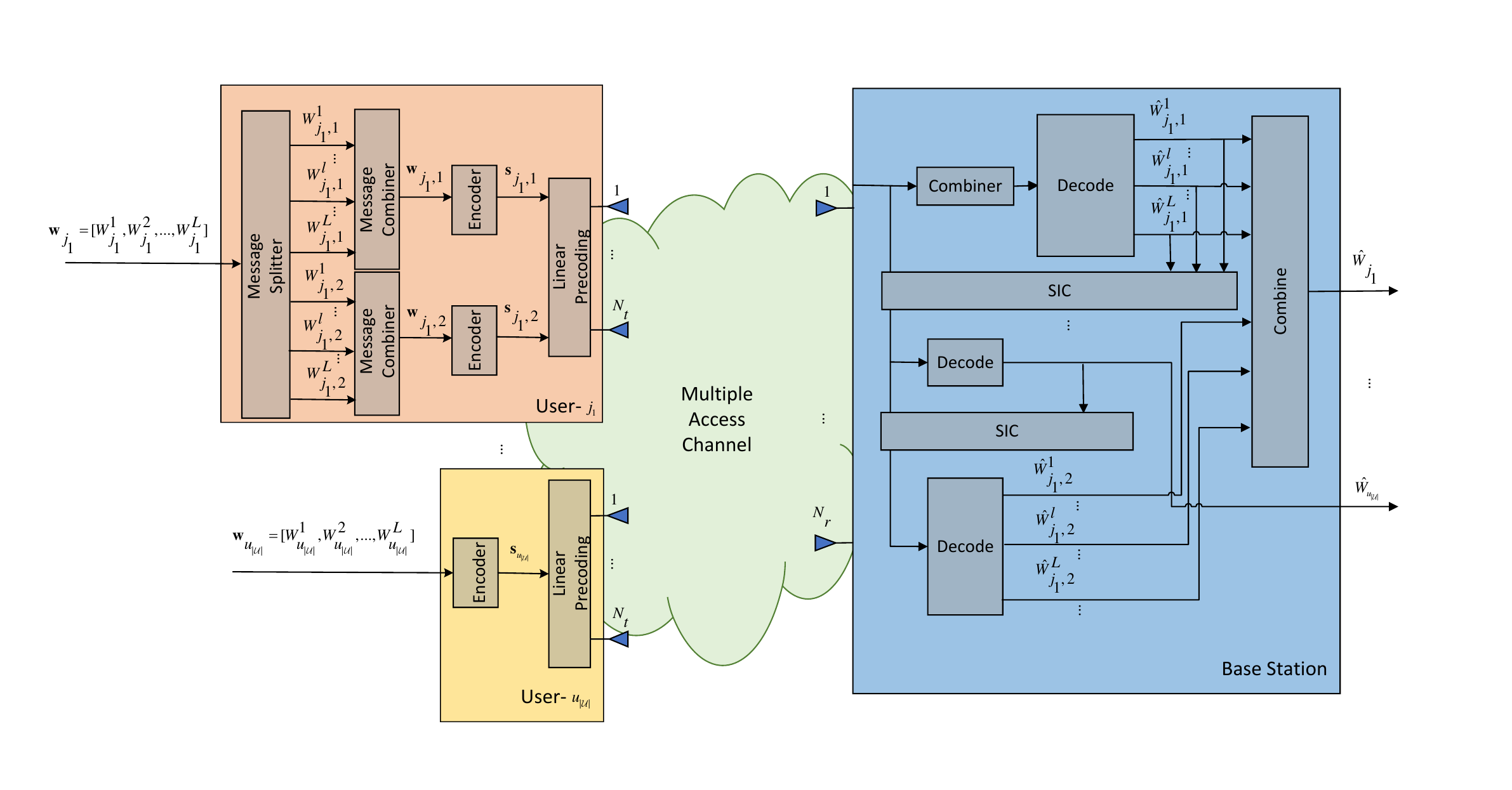}
    \caption{Transmission Model for uplink MU-MIMO adopting RSMA at the user side. Users are classified into splitting users (depicted in orange block) and non-splitting users (depicted in yellow block), with each user in the former group splitting their respective messages to experience different decoding priorities at the BS, leveraging the benefit of rate-splitting.}
    \label{Fig.1}
\end{figure*}
This section firstly models the uplink MIMO  RSMA framework, from message splitting strategies at each user's transmitters to the decoding strategy at the BS using SIC. Secondly, the uplink MIMO NOMA is also briefly introduced and explained as a subset of our RSMA scheme. We also express the achievable rates under both MIMO RSMA and NOMA scenarios with the practical FBL consideration, to facilitate future optimization and simulations.

\subsection{MIMO RSMA}\label{2.1}
Here we present the RSMA framework proposed for uplink MIMO channel. We consider a $K$-user uplink system with perfect CSIT and CSIR, where users indexed by $\mathcal{K}=\{1, 2,\ldots, K\}$ are equipped with $\textit{N}_{t}$ antennas. We assume $N_r$ antennas at the Base Station (BS), receiving messages from the $K$ users. The transmitted signal from user-$k, ~\textbf{x}_{k}\in\mathbb{C}^{N_t\times 1}, ~k \in \mathcal{K}$ is subject to a power constraint $\mathbb{E}\{\|\textbf{x}_{k}\|^{2}\}\leq P_{t}$. The signal is transmitted through a MIMO channel with $\textbf{H}_k\in\mathbb{C}^{N_r\times N_t}, ~k\in\mathcal{K}$ denoting the channel matrix between user-$k$ and BS. The transmission model of the proposed uplink MIMO with RSMA at the user side is illustrated in Fig. \ref{Fig.1} and explained as below. 
\subsubsection{Transmitter}
At the user side, let $L = \min\{N_t, N_r\}$ denote the number of message streams for user-$k$ and these messages are expressed as $\textbf{w}_k=\{W_{k}^{1}, W_{k}^{2},\ldots, W_{k}^{L}\}, ~k\in\mathcal{K}$. $\mathcal{J}\triangleq\{j_{1},\ldots, j_{|\mathcal{J}|}\}$ and $\mathcal{U}=\mathcal{K}\setminus\mathcal{J}\triangleq\{u_{1},\ldots, u_{|\mathcal{U}|}\}$ are defined as the splitting users set and the non-splitting users set, respectively. Each message in $\textbf{w}_{j_{l}}$ of user-$j_{l}$ is split into two parts $W_{j_{l}}^{i}=\{W_{j_{l},1}^{i}, W_{j_{l},2}^{i}\}, ~\forall i\in\{1,\ldots, L\}, ~\forall l\in\{1,\ldots, |\mathcal{J}|\}, ~j_{l}\in\mathcal{J}$. The first and second part of the messages for user-$j_{l}$ are combined together and denoted as $\textbf{w}_{j_{l},1}=\{W_{j_{l},1}^{1}, W_{j_{l},1}^{2},\ldots, W_{j_{l},1}^{L}\}$ and $\textbf{w}_{j_{l},2}=\{W_{j_{l},2}^{1}, W_{j_{l},2}^{2},\ldots, W_{j_{l},2}^{L}\}$, respectively. Separating each message into two parts leverages the freedom of varied decoding orders at the receiver. 
Then they are encoded into two symbol vectors $\textbf{s}_{j_{l},1}=[s_{j_{l},1}^{1}, s_{j_{l},1}^{2},\ldots, s_{j_{l},1}^{L}]^{T}\in\mathbb{C}^{L\times 1}$ and $\textbf{s}_{j_{l},2}=[s_{j_{l},2}^{1}, s_{j_{l},2}^{2},\ldots, s_{j_{l},2}^{L}]^{T}\in\mathbb{C}^{L\times 1}$, respectively. On the other hand, the messages of non-splitting user-$u_{q}$, $\textbf{w}_{u_{q}}$ are encoded into $\textbf{s}_{u_{q}}=[s_{u_{q}}^{1}, s_{u_{q}}^{2},\ldots, s_{u_{q}}^{L}]^{T}\in\mathbb{C}^{L\times 1}, ~\forall q\in\{1,\ldots, |\mathcal{U}|\}, ~u_{q}\in\mathcal{U}$. Therefore, by splitting $|\mathcal{J}|$ users' messages, there are $2|\mathcal{J}|+|\mathcal{U}|$ symbol vectors received at the BS in this $K$-user system, which are combined together into the vector $\textbf{s}=[\textbf{s}^{\mathcal{J}}, \textbf{s}^{\mathcal{U}}]^{T}$, where $\textbf{s}^{\mathcal{J}}=[\textbf{s}_{j_{1},1}, \textbf{s}_{j_{1},2},\ldots, \textbf{s}_{j_{|\mathcal{J}|},1}, \textbf{s}_{j_{|\mathcal{J}|},2}]^{T}$, and $\textbf{s}^{\mathcal{U}}=[\textbf{s}_{u_{1}},\ldots, \textbf{s}_{u_{|\mathcal{U}|}}]$, respectively. We use linear precoders $\textbf{P}_{j_{l}}=[\textbf{P}_{j_{l},1}, \textbf{P}_{j_{l},2}]$ to precode the symbol vectors of spliting user-$j_{l}$ with $tr(\textbf{P}_{j_{l}}\textbf{P}_{j_{l}}^{H})\leq P_{t}, ~j_{l}\in\mathcal{J}$. $\textbf{P}_{u_{q}}$ is used to precode the symbol vectors of non-spliting user-$u_{q}, ~tr(\textbf{P}_{u_{q}}\textbf{P}_{u_{q}})^{H}\leq P_{t}, ~u_{q}\in\mathcal{U}$, limited by the power budget $P_{t}$ at each user. Based on the above illustration, the received signal at the BS, $\textbf{y}\in\mathbb{C}^{N_{r}\times1}$, is given by
\begin{equation}\label{eq.1}
    \textbf{y} = \underbrace{\sum_{j_{l}\in\mathcal{J}}\textbf{H}_{j_{l}}\sum_{d=1}^{2}\textbf{P}_{j_{l},d}\textbf{s}_{j_{l},d}}_{\text{splitting users}}+\underbrace{\sum_{u_{q}\in\mathcal{U}}\textbf{H}_{u_{q}}\textbf{P}_{u_{q}}\textbf{s}_{u_{q}}}_{\text{non-splitting users}}+\textbf{n}, 
\end{equation}
where $\textbf{n}\sim\mathcal{CN}(0,\sigma_n^2\textbf{I}_{N_{r}})$ denotes the AWGN vector at the BS. Without loss of generality, we assume the noise variance is one, i.e., $\sigma_{n}^{2}=1$. Then, the transmit SNR is numerically equal to the transmit power.

\subsubsection{Receiver} At the BS side, the user signals are received and the SIC is employed. There are $(2|\mathcal{J}|+|\mathcal{U}|)!$ possible decoding orders. The optimal decoding order can be determined through exhaustive search, but its complexity grows exponentially with the number of users and data streams. To address this issue, we apply a low-complexity decoding order scheme proposed in our previous work \cite{xu2024max}, which involves separating the parts of splitting messages from users in $\mathcal{J}$ and arranging non-splitting users in $\mathcal{U}$ by the descending order of the channel gain. Assume that $\lVert\textbf{H}_{u_{j_{1}}}\rVert\geq\lVert\textbf{H}_{u_{j_{2}}}\rVert\geq\cdots\geq\lVert\textbf{H}_{u_{j_{|\mathcal{J}|}}}\rVert$ and $\lVert\textbf{H}_{u_{q_{1}}}\rVert\geq\lVert\textbf{H}_{u_{q_{2}}}\rVert\geq\cdots\geq\lVert\textbf{H}_{u_{q_{|\mathcal{U}|}}}\rVert$, the proposed decoding order is $\textbf{s}_{j_{1},1}\to\cdots\to \textbf{s}_{j_{|\mathcal{J}|},1}\to\textbf{s}_{u_{1}}\to\cdots\to\textbf{s}_{u_{|\mathcal{U}|}}\to\textbf{s}_{j_{1},2}\to\cdots\to\textbf{s}_{j_{|\mathcal{J}|},2}$.
We map this decoding order into $\bm{\pi}=[\textbf{s}_{\bm{\pi}_{1}},\ldots, \textbf{s}_{\bm{\pi}_{2|\mathcal{J}|+|\mathcal{U}|}}]$. These $2|\mathcal{J}|+|\mathcal{U}|$ symbol vectors are indexed by $\mathcal{M}=\{1,\ldots, 2|\mathcal{J}|+|\mathcal{U}|\}$. Therefore, the $m^{\text{th}}$ filtered received signal $\textbf{z}_{\bm{\pi}_{m}}, ~m\in\mathcal{M}$ before the $m^{\text{th}}$ stage of SIC process is 
\begin{equation} \label{eq.2}    
    \textbf{z}_{\bm{\pi}_{m}} = \underbrace{\textbf{G}_{\bm{\pi}_{m}}\textbf{H}_{\bm{\pi}_{m}}\textbf{P}_{\bm{\pi}_{m}}\textbf{s}_{\bm{\pi}_{m}}}_{\text{desired signal}} +\underbrace{\sum_{i=m+1}^{2|\mathcal{J}|+|\mathcal{U}|}\textbf{H}_{\bm{\pi}_{i}}\textbf{P}_{\bm{\pi}_{i}}\textbf{s}_{\bm{\pi}_{i}}+\textbf{G}_{\bm{\pi}_{m}}\textbf{n}}_{\text{interference plus noise}},  
 \end{equation}
where $\textbf{G}_{\bm{\pi}_{m}}\in\mathbb{C}^{L\times N_{r}}, ~m\in \mathcal{M}$ is the combiner which is used to detect symbol vector $\textbf{s}_{\bm{\pi}_{m}}$ and $\bm{\pi}_{m}$ refers to the $m^{\text{th}}$ symbol vector in the decoding order $\bm{\pi}$. This $m^{\text{th}}$ symbol vector $\textbf{s}_{\bm{\pi}_{m}}$
is decoded at the BS by treating the remaining $m+1^{\text{th}}$ to $2|\mathcal{J}|+|\mathcal{U}|^{\text{th}}$ symbol vectors as noise. Once $\textbf{s}_{\bm{\pi}_{m}}$ is successfully decoded, $\widehat{\textbf{w}}_{\bm{\pi}_{m}}$ is obtained, reconstructed and subtracted from the current received signal. 
\subsubsection{Rate with FBL constraints} Since we consider the FBL code here, the FBL achievable rate expression derived in \cite{polyanskiy2010channel} is applied. Then the rate of symbol vector $\textbf{s}_{\bm{\pi}_{m}}$ is written as
\begin{equation}\label{eq.3}
    R_{\bm{\pi}_{m}}=\sum_{a=1}^{L}\log_{2}(1+\gamma_{\bm{\pi}_{m}}^{a})-\frac{B}{N}\sqrt{V_{\gamma_{\bm{\pi}_{m}}^{a}}},
\end{equation}
where $\gamma_{\bm{\pi}_{m}}^{a}$ denotes the Signal-to-Interference-Noise-Ratio (SINR) of stream $a$ of symbol vector $\textbf{s}_{\bm{\pi}_{m}}$ 
\begin{equation} \label{eq.4}
    \gamma_{\bm{\pi}_{m}}^{a}  = \frac{|\textbf{g}_{\bm{\pi}_{m}}^{a}\textbf{H}_{\bm{\pi}_{m}}\textbf{p}_{\bm{\pi}_{m}}^{a}|^2}{I_{l,\bm{\pi}_{m}}^{a}+I_{c,\bm{\pi}_{m}}^{a}+\lVert \textbf{g}_{\bm{\pi}_{m}}^{a} \rVert\sigma_{n}^{2}},  
\end{equation}
with $\textbf{p}_{\bm{\pi}_{m}}^{a}$ (resp. $\textbf{g}_{\bm{\pi}_{m}}^{a}$) the precoder (resp. combiner) attached to stream $a$ of symbol vector $\textbf{s}_{\bm{\pi}_{m}}$ and is the $a^{\text{th}}$ column (resp. row) of $\textbf{P}_{\bm{\pi}_{m}}$ (resp. $\textbf{G}_{\bm{\pi}_{m}}$). $I_{l,\bm{\pi}_{m}}^{a}$ is the inter-stream interference and $I_{c,\bm{\pi}_{m}}^{a}$ is the interference from other symbol vectors suffered by stream $a$ of symbol vector $\textbf{s}_{\bm{\pi}_{m}}$ which are given as
\begin{subequations}
    \begin{align}
        I_{l,\bm{\pi}_{m}}^{a} & = \sum_{i=1, i\neq a}^{L}|\textbf{g}_{\bm{\pi}_{m}}^{a}\textbf{H}_{\bm{\pi}_{m}}\textbf{p}_{\bm{\pi}_{m}}^{i}|^2  \label{eq.5a}\\
        I_{c,\bm{\pi}_{m}}^{a} & = \sum_{j=m+1}^{2|\mathcal{J}|+|\mathcal{U}|}\sum_{i=1}^{L}|\textbf{g}_{\bm{\pi}_{m}}^{a}\textbf{H}_{\bm{\pi}_{j}}\textbf{p}_{\bm{\pi}_{j}}^{i}|^{2}. \label{eq.5b}
    \end{align}
\end{subequations}
As for the penalty part brought by FBL, $B=Q^{-1}(\epsilon)\log_2(e)$, $\epsilon$ is the error probability and $Q$ is the Q-function. \footnote{Q-function is the tail distribution function of the standard normal distribution. Normally, Q-function is defined as: $Q(x)=\frac{1}{\sqrt{2\pi}} \int_x^{\infty}e^{-\frac{u^2}{2}}du.$} $N$ is the blocklength and $V = \biggl(1-(1+\gamma_{\bm{\pi}_{m}}^{a})^{-2}\biggl)$ is the channel dispersion.
The total achievable rates of splitting users can be calculated by summing the transmission rates for the two corresponding split streams, e.g., the rate of user-$j_{1}$ in $\mathcal{J}$ equals $R_{\bm{\pi}_{1}}+R_{\bm{\pi}_{1+K}}$ 
and the rates of non-splitting users such as user-$u_{1}$ in $\mathcal{U}$ equals $R_{\bm{\pi}_{|\mathcal{J}|+1}}$ individually. 

\subsection{MIMO NOMA}\label{2.2}
The system model of MIMO NOMA is a particular instance of the model specified in Sec.\ref{2.1}. By grouping all users into the non-splitting user set, MIMO RSMA boils down to MIMO NOMA. Let $L= \min\{N_t,N_r\}$ denote the number of message streams for user-$k$ and they are expressed as $\textbf{w}_k=\{W_{k}^{1}, W_{k}^{2},\ldots, W_{k}^{L}\}, ~k\in\mathcal{K}$ and then are encoded into $\textbf{s}_{k}=[s_{k}^{1}, s_{k}^{2},\ldots, s_{k}^{L}]^{T}\in\mathbb{C}^{L\times 1}$. $\textbf{P}_{k}$ is used to precode the symbol vector of user-$k, \textit{tr}(\textbf{P}_{k}(\textbf{P}_{k})^{H})\leq P_{t}$. The received signal at the BS, $\textbf{y}\in\mathbb{C}^{N_{r}\times1}$, is given by
\begin{equation} \label{eq.6}
    \textbf{y} = \sum_{k\in\mathcal{K}}\textbf{H}_{k}\textbf{P}_{k}\textbf{s}_{k}+\textbf{n}.
\end{equation}
The decoding order is arranged according to the descending order of the channel gain and we assume that $\lVert\textbf{H}_{1}\rVert\geq\lVert\textbf{H}_{2}\rVert\geq\cdots\geq\lVert\textbf{H}_{K}\rVert$. The proposed decoding order is $\textbf{s}_{1}\to\textbf{s}_{2}\to\cdots\to\textbf{s}_{K-1}\to\textbf{s}_{K}$. Then the SINR of stream $a$ of symbol vector $\textbf{s}_{k}, ~k\in\mathcal{K}$ is written as
\begin{equation} \label{eq.7}
    \gamma_{k}^{a}  = \frac{|\textbf{g}_{k}^{a}\textbf{H}_{k}\textbf{p}_{k}^{a}|^2}{I_{l,k}^{a}+I_{c,k}^{a}+\lVert \textbf{g}_{k}^{a} \rVert\sigma_{n}^{2}},  
\end{equation} 
where $\textbf{p}_{k}^{a}$ and $\textbf{g}_{k}^{a}$ is $a^{\text{th}}$ column of $\textbf{P}_{k}$ and  $a^{\text{th}}$ row of $\textbf{G}_{k}$ of stream $a$ in symbol vector $\textbf{s}_{k}$, respectively. $I_{l,k}^{a}$ and $I_{c,k}^{a}$ is the inter-stream interference and interference from other symbol vectors experienced by stream $a$ of symbol vector $\textbf{s}_{k}$ which are given as
\begin{subequations}
    \begin{align}
        I_{l,k}^{a} & = \sum_{i=1, i\neq a}^{L}|\textbf{g}_{k}^{a}\textbf{H}_{k}\textbf{p}_{k}^{i}|^2  \label{eq.8a}\\
        I_{c,k}^{a} & = \sum_{j=k+1}^{2|\mathcal{J}|+|\mathcal{U}|}\sum_{i=1}^{L}|\textbf{g}_{k}^{a}\textbf{H}_{j}\textbf{p}_{j}^{i}|^{2}. \label{eq.8b}
    \end{align}
\end{subequations}
Then the FBL achievable rate expression of symbol vector $\textbf{s}_{k}$ is written as
\begin{equation}\label{eq.9}
    R_{k}=\sum_{a=1}^{L}\log_{2}(1+\gamma_{k}^{a})-\frac{B}{N}\sqrt{V_{\gamma_{k}^{a}}}.
\end{equation}
For simplicity, MIMO RSMA and MIMO NOMA will be referred to as RSMA and NOMA for the rest of the paper respectively.


\begin{figure}[t]
    \centering
    \includegraphics[scale=0.5]{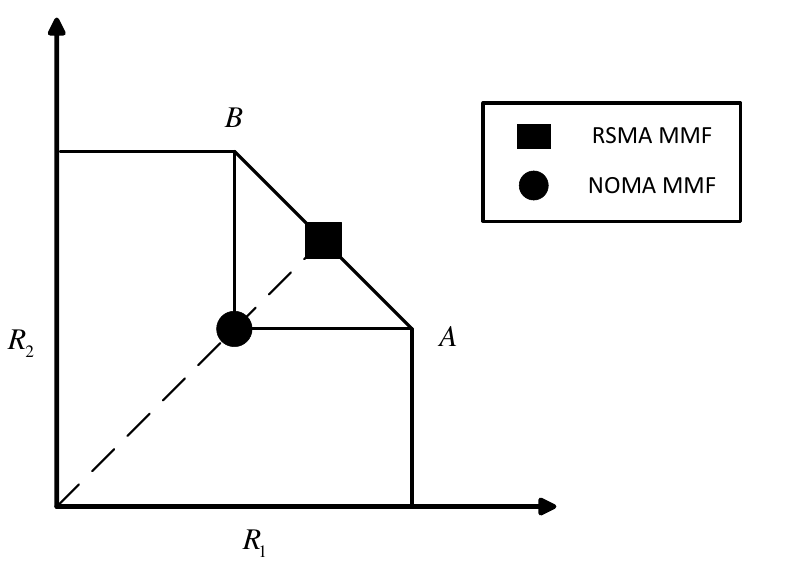}
    \caption{The schematic illustration of the two-user rate-region. The highlighted points (RSMA marked by a square and NOMA marked by a circle) represent the best fairness achieved by RSMA and NOMA respectively.}
    \label{Fig.2}
\end{figure} 

\remark Here, we take a two-user example to illustrate the benefits of RSMA over NOMA regarding the MMF metric. The capacity region for RSMA and NOMA with random transmit power is illustrated in Fig. \ref{Fig.2}. The capacity region always has the shape of a pentagon. The two corner points, $A$ and $B$, correspond to the maximum rates achieved by two SIC decoding orders: A is achieved by decoding user-1 first and B is achieved by decoding user-2 first. The points on the segment $AB$ can be achieved with time-sharing with two reverse decoding orders for NOMA. However, RSMA can directly achieve the points between two corner points without time-sharing. Therefore, without considering time-sharing, MMF achieved by RSMA is higher than that of NOMA which is represented by the black square in Fig. \ref{Fig.2} and MMF of NOMA is denoted by the black circle.

\section{Problem Formulation and Optimization} \label{3}
In this section, we formulate the optimization problem to design the precoders and combiners to facilitate the RSMA performance, with the primary purpose of guaranteeing communication fairness between users. We develop an effective algorithm to solve the problem and provide the numerical complexity analysis for the proposed algorithm.

\subsection{Problem Formulation}
In this section, we formulate an optimization problem with respect to both precoders and combiners for the proposed system model. We emphasize the user fairness issue in the objective function. Hence, the MMF problem for the $K$-user uplink RSMA system is formulated as
\begin{subequations} \label{Prob.10}
    \begin{align}
        \max_{\textbf{P}_{j_{l}}, \textbf{P}_{u_{q}}, \textbf{G}} \quad \min_{k}{R_{k}} \\
        \mbox{s.t.} \quad
        & \text{tr}(\textbf{P}_{j_{l}}\textbf{P}_{j_{l}}^{H})\leq P_{t}, ~j_{l}\in \mathcal{J}, \\
        & \text{tr}(\textbf{P}_{u_{q}}\textbf{P}_{u_{q}}^{H})\leq P_{t}, ~u_{q}\in\mathcal{U},
    \end{align} 
\end{subequations}
where $R_{k}$ is the rate of user-$k$ in $\mathcal{K}$ and $\textbf{G}$ is the combiner. We assume the maximum transmit power $P_{t}$ is equal for all users. We proposed a SCA-based algorithm to solve Problem \eqref{Prob.10} because the rate and SINR equations are not convex. This algorithm involves adding slack variables and progressively approximating the non-convex expressions using their first-order Taylor approximation.

First an auxiliary variable $t$ is introduced and for the given decoding order $\bm{\pi}$, Problem \eqref{Prob.10} is transferred into
\begin{subequations} \label{Prob.11}
    \begin{align}
        & \max_{\textbf{P}_{\bm{\pi}}, \textbf{G}_{\bm{\pi}}, t} \quad t \\
        \mbox{s.t.} \quad
        & R_{l}+R_{l+K}\geq t, ~l\in\{1,\ldots, |\mathcal{J}|\}\label{eq.11b}\\
        & R_{o}\geq t, ~o\in\{|\mathcal{J}|,\ldots, K\} \label{eq.11c}\\   
        & \text{tr}(\textbf{P}_{l}\textbf{P}_{l}^{H})+\text{tr}(\textbf{P}_{l+K}\textbf{P}_{l+K}^{H})\leq P_{t}, ~l\in\{1,\ldots, |\mathcal{J}|\} \label{eq.11d} \\
        & \text{tr}(\textbf{P}_{o}\textbf{P}_{o}^{H})\leq P_{t}, ~o\in\{|\mathcal{J}|,\ldots, K\}, \label{eq.11e}
    \end{align} 
\end{subequations} 
where $\textbf{P}_{\bm{\pi}}=[\textbf{P}_{j_{1},1}, ..., \textbf{P}_{j_{|\mathcal{J}|},1}, \textbf{P}_{u_{1}}, ..., \textbf{P}_{u_{|\mathcal{U}|}}, \textbf{P}_{j_{1},2}, ..., \textbf{P}_{j_{|\mathcal{J}|},2}]$ and $\textbf{G}_{\bm{\pi}}=[\textbf{G}_{j_{1},1}, ..., \textbf{G}_{j_{|\mathcal{J}|},1}, \textbf{G}_{u_{1}}, ..., \textbf{G}_{u_{|\mathcal{U}|}}, \textbf{G}_{j_{1},2}, ..., \textbf{G}_{j_{|\mathcal{J}|},2}]$ are the precoders and combiners of each symbol vector corresponding to the decoding order $\bm{\pi}$, respectively. \eqref{eq.11b} and \eqref{eq.11c} refer to the maximization of minimum user rate constraints of every splitting user and non-splitting. In the decoding order $\bm{\pi}$, the symbol vectors $\textbf{s}_{j_{l},1}$ and $\textbf{s}_{j_{l},2}$ are mapped to $\textbf{s}_{l}$ and $\textbf{s}_{l+K}, ~l\in\{1, \ldots, |\mathcal{J}|\}$ and $\textbf{s}_{u_{q}}$ is mapped to $s_{o}, ~o\in\{|\mathcal{J}|,\ldots, K\}$ with both $\{1, \ldots, |\mathcal{J}|\}$ and $\{|\mathcal{J}|, \ldots, K\}$ the subsets of $\mathcal{M}$. We further introduce slack variables $\bm{\rho}_{\bm{\pi}}=[\bm{\rho}_{1}, \ldots, \bm{\rho}_{2|J|+|\mathcal{U}|}]$ where $\bm{\rho}_{m}=[\rho_{m}^{1},\ldots, \rho_{m}^{L}]^{T}, ~m\in\mathcal{M}$. Then Problem \eqref{Prob.11} is transformed into
\begin{subequations}\label{Prob.12}
    \begin{align}
        & \max_{\textbf{P}_{\bm{\pi}}, \textbf{G}_{\bm{\pi}}, \bm{\rho}_{\bm{\pi}}, t} \quad t \\
        \mbox{s.t.} \quad
        & \sum_{a=1}^{L}\left(\log_2(1+\rho_{l}^{a})-\frac{B}{\sqrt{N}}\sqrt{V_{\rho_{l}^{a}}}+\log_2(1+\rho_{l+K}^{a})\right. \nonumber \\
        & \left.-\frac{B}{\sqrt{N}}\sqrt{V_{\rho_{l+K}^{a}}}\right)\geq t, ~l\in(1, \ldots, |\mathcal{J}|) \label{eq.12b} \\
        & \sum_{a=1}^{L}\left(\log_2(1+\rho_{o}^{a})-\frac{B}{\sqrt{N}}\sqrt{V_{\rho_{o}}}\right)\geq t, ~o\in(|\mathcal{J}|,\ldots, K) \label{eq.12c} \\
        & \frac{|\textbf{g}_{\bm{\pi}_{m}}^{a}\textbf{H}_{\bm{\pi}_{m}}\textbf{p}_{\bm{\pi}_{m}}^{a}|^2}{I_{l,\bm{\pi}_{m}}^{a}+I_{c,\bm{\pi}_{m}}^{a}+\lVert \textbf{g}_{\bm{\pi}_{m}}^{a} \rVert\sigma_{n}^{2}}\geq\rho_{m}^{a}, ~m\in\mathcal{M}, \nonumber \\
        & a\in\{1,\ldots, L_{\bm{\pi}_{m}}\}\label{eq.12d} \\
        & \eqref{eq.11d}, \eqref{eq.11e}. \nonumber 
    \end{align} 
\end{subequations}
However, Problem \eqref{Prob.12} is still non-convex due to the term $\sqrt{V_{\rho_{l}^{a}}}$, $\sqrt{V_{\rho_{l}^{a}}}$ and $\sqrt{V_{\rho_{l+K}^{a}}}$ in constraint \eqref{eq.12b} and \eqref{eq.12c}. Thus we approximate these three non-convex terms by their first-order Taylor approximation around the point $\left(\bm{\rho}_{\bm{\pi}}^{[n]}\right)$ at iteration $n$, which are expressed as 
\begin{subequations}
    \begin{align}
        \sqrt{V_{\rho_{l}^{a}}} & \leq \sqrt{1-\left(1+(\rho_{l}^{a})^{[n]}\right)^{-2}}+\left(\rho_{l}^{a}-(\rho_{l}^{a})^{[n]}\right) \nonumber \\
        & \left(1+(\rho_{l}^{a})^{[n]}\right)^{-3}\left(1-(1+(\rho_{l}^{a})^{[n]})^{-2}\right)^{-\frac{1}{2}} \nonumber\\
        & \triangleq \left(\Phi_{l}^{a}\right)^{[n]} \\
        \sqrt{V_{\rho_{l+K}^{a}}} & \leq \sqrt{1-\left(1+(\rho_{l+K}^{a})^{[n]}\right)^{-2}}+\left(\rho_{l+K}^{a}-(\rho_{l+K}^{a})^{[n]}\right) \nonumber \\
        & \left(1+(\rho_{l+K}^{a})^{[n]}\right)^{-3}\left(1-(1+(\rho_{l+K}^{a})^{[n]})^{-2}\right)^{-\frac{1}{2}} \nonumber\\
        & \triangleq \left(\Phi_{l+K}^{a}\right)^{[n]} \\
        \sqrt{V_{\rho_{o}}} & \leq \sqrt{1-\left(1+(\rho_{o}^{a}\right)^{[n]})^{-2}}+\left(\rho_{o}^{a}-(\rho_{o}^{a})^{[n]}\right) \nonumber \\
        &\left(1+(\rho_{o}^{a})^{[n]}\right)^{-3}\left(1-(1+(\rho_{o}^{a})^{[n]})^{-2}\right)^{-\frac{1}{2}} \nonumber\\
        & \triangleq \left(\Phi_{o}^{a}\right)^{[n]}.
    \end{align}
\end{subequations} 
Thus, Problem \eqref{Prob.12} is transformed into 
\begin{subequations}\label{Prob.14}
    \begin{align}
        & \max_{\textbf{P}_{\bm{\pi}}, \textbf{G}_{\bm{\pi}}, \bm{\rho}_{\bm{\pi}}, t} \quad t \\
        \mbox{s.t.} \quad
        & \sum_{a=1}^{L}\left(\log_2(1+\rho_{l}^{a})-\frac{B}{\sqrt{N}}(\Phi_{l}^{a})^{[n]}+\log_2(1+\rho_{l+K}^{a}) \right.\nonumber \\
        & \left.-\frac{B}{\sqrt{N}}(\Phi_{l+K}^{a})^{[n]}\right)\geq t, ~l\in(1,\ldots, |\mathcal{J}|) \label{eq.14b} \\
        & \sum_{a=1}^{L}\left(\log_2(1+\rho_{o}^{a})-\frac{B}{\sqrt{N}}(\Phi_{o}^{a})^{[n]}\right)\geq t, ~o\in(|\mathcal{J}|,\ldots, K) \label{eq.14c} \\
        & \frac{|\textbf{g}_{\bm{\pi}_{m}}^{a}\textbf{H}_{\bm{\pi}_{m}}\textbf{p}_{\bm{\pi}_{m}}^{a}|^2}{I_{l,\bm{\pi}_{m}}^{a}+I_{c,\bm{\pi}_{m}}^{a}+\lVert \textbf{g}_{\bm{\pi}_{m}}^{a} \rVert\sigma_{n}^{2}}\geq\rho_{m}^{a}, ~m\in\mathcal{M}, \nonumber \\
        & a\in\{1,\ldots, L_{\bm{\pi}_{m}}\}\label{eq.14d} \\
        & \eqref{eq.11d}, \eqref{eq.11e}. \nonumber
    \end{align} 
\end{subequations}
Since precoder $\textbf{p}_{\bm{\pi}_{m}}^{a}$ and the combiner $\textbf{g}_{\bm{\pi}_{m}}^{a}$ are highly coupled in \eqref{eq.14d}, we decompose Problem \eqref{Prob.14} into the two subproblems of precoders optimization and combiners optimization, which can be efficiently solved by the SCA technique as described in the following.

\subsection{Precoders Optimization} \label{3.A}
Under any given feasible combiner $\textbf{G}_{\bm{\pi}_{m}}, ~m\in\mathcal{M}$ with  $\textbf{g}_{\bm{\pi}_{m}}^{a}$ is the combiner for stream $a$ of symbol vector $\textbf{s}_{\bm{\pi}}$, Problem \eqref{Prob.14} is rewritten as
\begin{subequations}\label{Prob.15}
    \begin{align}
        & \max_{\textbf{P}_{\bm{\pi}}, \bm{\rho}_{\bm{\pi}}, t} \quad t \\
        \mbox{s.t.} \quad
        & \eqref{eq.14b}, \eqref{eq.14c}, \eqref{eq.14d}, \eqref{eq.11d}, \eqref{eq.11e}. \nonumber
    \end{align} 
\end{subequations}
With fixed combiner $\textbf{g}_{\bm{\pi}_{m}}^{a}$, constraint \eqref{eq.14d} can be rewritten as
\begin{equation} \label{eq.16}
    \frac{|\textbf{g}_{\bm{\pi}_{m}}^{a}\textbf{H}_{\bm{\pi}_{m}}\textbf{p}_{\bm{\pi}_{m}}^{a}|^2}{\rho_{m}^{a}}\geq I_{l,\bm{\pi}_{m}}^{a}+I_{c,\bm{\pi}_{m}}^{a}+\lVert\textbf{g}_{\bm{\pi}_{m}}^{a}\rVert\sigma_{n}^{2}.
\end{equation}
Since \eqref{eq.16} is not convex, for the Left-Hand Side (LHS) of \eqref{eq.16}, we use the first-order Taylor approximation to approximate it around the point $\left(\textbf{P}_{\bm{\pi}}^{[n]}, \bm{\rho}_{\bm{\pi}}^{[n]}\right)$ at iteration $n$ and it can be expressed as
\begin{equation} \label{eq.17}
\begin{split}
    \frac{|\textbf{g}_{\bm{\pi}_{m}}^{a}\textbf{H}_{\bm{\pi}_{m}}\textbf{p}_{\bm{\pi}_{m}}^{a}|^2}{\rho_{m}^{a}} & \geq\frac{2\textit{R}\{\textbf{g}_{\bm{\pi}_{m}}^{a}\textbf{H}_{\bm{\pi}_{m}}\left((\textbf{p}_{\bm{\pi}_{m}}^{a})^{[n]})^{H}(\textbf{H}_{\bm{\pi}_{m}}\right)^{H}(\textbf{g}_{\bm{\pi}_{m}}^{a})^{H}\}}{(\bm{\rho}_{\bm{\pi}_{m}}^{a})^{[n]}} \\
    & -\frac{\rho_{m}^{a}|\textbf{g}_{\bm{\pi}_{m}}^{a}\textbf{H}_{\bm{\pi}_{m}}(\textbf{p}_{\bm{\pi}_{m}}^{a})^{[n]}|^2}{\left((\rho_{m}^{a})^{[n]}\right)^{2}}.
\end{split}
\end{equation}
Then, constraint \eqref{eq.14d} is rewritten as
\begin{equation} \label{eq.18}
\begin{split}
    I_{l,\bm{\pi}_{m}}^{a} & +I_{c,\bm{\pi}_{m}}^{a}+\lVert\textbf{g}_{\bm{\pi}_{m}}^{a}\rVert\sigma_{n}^{2}+\frac{\rho_{m}^{a}|\textbf{g}_{\bm{\pi}_{m}}^{a}\textbf{H}_{\bm{\pi}_{m}}(\textbf{p}_{\bm{\pi}_{m}}^{a})^{[n]}|^2}{\left((\rho_{m}^{a})^{[n]}\right)^{2}} \\
    & -\frac{2\textit{R}\{\textbf{g}_{\bm{\pi}_{m}}^{a}\textbf{H}_{\bm{\pi}_{m}}\left((\textbf{p}_{\bm{\pi}_{m}}^{a})^{[n]}\right)^{H}(\textbf{H}_{\bm{\pi}_{m}})^{H}(\textbf{g}_{\bm{\pi}_{m}}^{a})^{H}\}}{(\rho_{m}^{a})^{[n]}}\leq 0.
\end{split}
\end{equation}
Therefore, with a given combiner $\textbf{G}_{\bm{\pi}}$, the original Problem \eqref{Prob.15} is transformed into 
\begin{subequations}\label{Prob.19}
    \begin{align}
        \max_{\textbf{P}_{\bm{\pi}}, \bm{\rho}, t} \quad t \\
        \mbox{s.t.} \quad
        & \eqref{eq.14b},\eqref{eq.14c}, \eqref{eq.18}, \eqref{eq.11d}, \eqref{eq.11e}. \nonumber
    \end{align} 
\end{subequations}
Note that the objective value obtained from Problem \eqref{Prob.19} provides a lower bound on that of
Problem \eqref{Prob.15} due to the replacement of the lower bounds in \eqref{eq.17}.

\subsection{Combiners Optimization} \label{3.2}
Similar to precoders optimization, for any given feasible precoder $\textbf{P}_{\bm{\pi}_{m}}, ~m\in\mathcal{M}$ where $\textbf{p}_{\bm{\pi}_{m}}^{a}$ is the precoder of stream $a$ of symbol vector $\textbf{s}_{\bm{\pi}}$,
Problem \eqref{Prob.14} is rewritten as
\begin{subequations}\label{Prob.20}
    \begin{align}
        \max_{\textbf{G}_{\bm{\pi}}, \bm{\rho}_{\bm{\pi}}, t} \quad t \\
        \mbox{s.t.} \quad
        & \eqref{eq.14b}, \eqref{eq.14c}, \eqref{eq.14d}. \nonumber
    \end{align} 
\end{subequations}
It is still the non-convexity of constraint \eqref{eq.14d} that causes the problem hard to solve. By applying the same method used in Sec. \ref{3.A}, we approximate constraint \eqref{eq.14d} around the point $\left(\textbf{G}_{\bm{\pi}}^{[n]}, \bm{\rho}_{\bm{\pi}}^{[n]}\right)$ at iteration $n$ and it can be written as
\begin{equation} \label{eq.21}
    \begin{split}
    I_{l,\bm{\pi}_{m}}^{a} & +I_{c,\bm{\pi}_{m}}^{a}+\lVert\textbf{g}_{\bm{\pi}_{m}}^{a}\rVert\sigma_{n}^{2}+\frac{\rho_{m}^{a}|(\textbf{g}_{\bm{\pi}_{m}}^{a})^{[n]}\textbf{H}_{\bm{\pi}_{m}}\textbf{p}_{\bm{\pi}_{m}}^{a}|^2}{\left((\rho_{m}^{a})^{[n]}\right)^{2}} \\
    & -\frac{2\textit{R}\{\textbf{g}_{\bm{\pi}_{m}}^{a}\textbf{H}_{\bm{\pi}_{m}}(\textbf{p}_{\bm{\pi}_{m}}^{a})^{H}(\textbf{H}_{\bm{\pi}_{m}})^{H}\left((\textbf{g}_{\bm{\pi}_{m}}^{a})^{[n]}\right)^{H}\}}{(\rho_{m}^{a})^{[n]}}\leq 0.
    \end{split}    
\end{equation}
Therefore, with a given combiner $\textbf{p}_{\bm{\pi}_{m}}^{a}$, the original Problem \eqref{Prob.19} is transformed into 
\begin{subequations}\label{Prob.22}
    \begin{align}
        \max_{\textbf{G}_{\bm{\pi}}, \bm{\rho}_{\bm{\pi}}, t} \quad t \\
        \mbox{s.t.} \quad
        & \eqref{eq.14b}, \eqref{eq.14c}, \eqref{eq.21}.
\nonumber
    \end{align} 
\end{subequations}
Similarly, the objective value obtained from Problem \eqref{Prob.22} serves as a lower bound on that of Problem \eqref{Prob.20}. Now, it is easy to verify that both Problem \eqref{Prob.19} and \eqref{Prob.22} are convex problems, which can be efficiently solved via standard convex problem solvers such as CVX \cite{grant2009cvx}.

\begin{algorithm} [t] \label{Alg.1}
\caption{Proposed SCA-based AO algorithm for solving Problem \eqref{Prob.11}}
\LinesNumbered
\SetKwInput{kwInit}{Initialize}
\kwInit{$n\gets0,t^{[n]}\gets0, \text{and feasible} 
 \mathbf{P}_{\bm{\pi}}^{[n]}, \mathbf{G}_{\bm{\pi}}^{[n]}$;}
\Repeat{$|t^{[n]}-t^{[n-1]}|\leq \tau$}
 {$n\gets n+1$; \\
 Find optimal $\mathbf{P}_{\bm{\pi}}^{[n]}$ by solving Problem \eqref{Prob.19} for given $\mathbf{G}_{\bm{\pi}}^{[n]-1}$; \\
 Find optimal $\mathbf{G}_{\bm{\pi}}^{[n]}$ by solving Problem \eqref{Prob.22} for given $\mathbf{P}_{\bm{\pi}}^{[n]}$; \\
 Update $t^{[n]}\gets t^*, \mathbf{P}_{\bm{\pi}}^{[n]}\gets \mathbf{P}_{\bm{\pi}}^*, \mathbf{G}_{\bm{\pi}}^{[n]}\gets \mathbf{G}_{\bm{\pi}}^*$;}
\end{algorithm}

\subsection{Proposed Algorithm, Convergence and Complexity}
According to the above two subproblems, we apply the AO to solve Problem \eqref{Prob.11} by utilizing the SCA-based algorithm. Specifically, the precoder $\textbf{P}_{\bm{\pi}}$ and the combiner $\textbf{G}_{\bm{\pi}}, ~m\in\mathcal{M}$ are alternately optimized by solving Problem \eqref{Prob.19} and \eqref{Prob.22}. The details of the proposed algorithm are summarized in \textbf{Algorithm 1}. At each iteration of Algorithm 1, the precoders and combiners are updated by solving Second Order Cone Programming (SOCP) problems. Each SOCP is solved by using interior-point method with the computational complexity of $\mathcal{O}\left([X]^{3.5}\right)$, where $X$ is the total number of variables in the corresponding SOCP problem. Although an additional variable $\bm{\rho}$ is introduced for approximating convex problems, the main complexity still comes from the precoders and combiners design. With given combiners, the number of variables of Problem \eqref{Prob.19} is given by $X_{\text{precoder}}=\left(2|\mathcal{J}|+|\mathcal{U}|\right)N_{t}$. Similarly, the number of variables of Problem \eqref{Prob.22} is given by $X_{\text{combiner}}=\left(2|\mathcal{J}|+|\mathcal{U}|\right)N_{r}$. The total number of iterations required for the convergence is $\mathcal{O}\left(\log(\tau^{-1})\right)$, where $\tau$ is the convergence tolerance of Algorithm 1. Therefore, the total computation complexity of Algorithm 1 is $\mathcal{O}\left((2|\mathcal{J}|+|\mathcal{U}|)(N_{t}+N_{r})^{3.5}\log(\tau^{-1})\right)$. 

Next we demonstrate the convergence of Algorithm 1. Define $t\left(\textbf{P}_{\bm{\pi}}^{[n]}, \textbf{G}_{\bm{\pi}}^{[n]}\right)$ as the objective value at the $n^{\text{th}}$ iteration. First, from Problem \eqref{Prob.19} with a given $\textbf{G}_{\bm{\pi}}$ in step 3 of Algorithm 1, we know
\begin{equation} \label{eq.23}
    \begin{split}
        t\left(\textbf{P}_{\bm{\pi}}^{[n-1]}, \textbf{G}_{\bm{\pi}}^{[n-1]}\right) & \overset{a}{=} t_{\textbf{P}}\left(\textbf{P}_{\bm{\pi}}^{[n-1]}, \textbf{G}_{\bm{\pi}}^{[n-1]}\right) \\
        & \overset{b}{\leq}t_{\textbf{P}}\left(\textbf{P}_{\bm{\pi}}^{[n]}, \textbf{G}_{\bm{\pi}}^{[n-1]}\right) \\
        & \overset{c}{\leq}t\left(\textbf{P}_{\bm{\pi}}^{[n]}, \textbf{G}_{\bm{\pi}}^{[n]}\right),
    \end{split}
\end{equation}
where $t_{\textbf{P}}$ represents the objective value of Problem \eqref{Prob.19}. $\textit{a}$ holds since the first-order Taylor approximations are tight at the given point $\left(\textbf{P}_{\bm{\pi}}^{[n-1]}, \bm{\rho}^{[n-1]}\right)$. Since the solution of the approximated Problem \eqref{Prob.19} at the $n-1^{\text{th}}$ iteration is a feasible point for Problem \eqref{Prob.19} at the $n^{\text{th}}$ iteration, it can be solved successfully. Moreover, the objective function is bounded by the transmit power constraints, then \textit{b} holds. \textit{c} is due to the objective value of \eqref{Prob.19} being the lower bound of \eqref{Prob.15}.

Similarly, from Problem \eqref{Prob.22} with a given $\textbf{P}_{\bm{\pi}}$ in step 4 of Algorithm 1, we know
\begin{equation} \label{eq.24}
    \begin{split}
        t\left(\textbf{P}_{\bm{\pi}}^{[n]}, \textbf{G}_{\bm{\pi}}^{[n]}\right) & \overset{a}{=} t_{\textbf{G}}\left(\textbf{P}_{\bm{\pi}}^{[n]}, \textbf{G}_{\bm{\pi}}^{[n]}\right) \\
        & \overset{b}{\leq}t_{\textbf{G}}\left(\textbf{P}_{\bm{\pi}}^{[n]}, \textbf{G}_{\bm{\pi}}^{[n-1]}\right) \\
        & \overset{c}{\leq}t\left(\textbf{P}_{\bm{\pi}}^{[n]}, \textbf{G}_{\bm{\pi}}^{[n]}\right),
    \end{split}
\end{equation}
where $t_{\textbf{G}}$ represents the objective value of Problem \eqref{Prob.22}. $\textit{a}$, $\textit{b}$ and $\textit{c}$ hold for the same reasons as we described previously.
Thus, we can obtain that
\begin{equation} \label{eq.25}
    t\left(\textbf{P}_{\bm{\pi}}^{[n-1]}, \textbf{G}_{\bm{\pi}}^{[n-1]}\right)\leq t\left(\textbf{P}_{\bm{\pi}}^{[n]}, \textbf{G}_{\bm{\pi}}^{[n]}\right),
\end{equation}
which proves that the Algorithm 1 generates a non-decreasing sequence of objective values.

\section{Physical Layer Design} \label{4}
\begin{figure*}[t]
    \centering
    \includegraphics[scale=0.5]{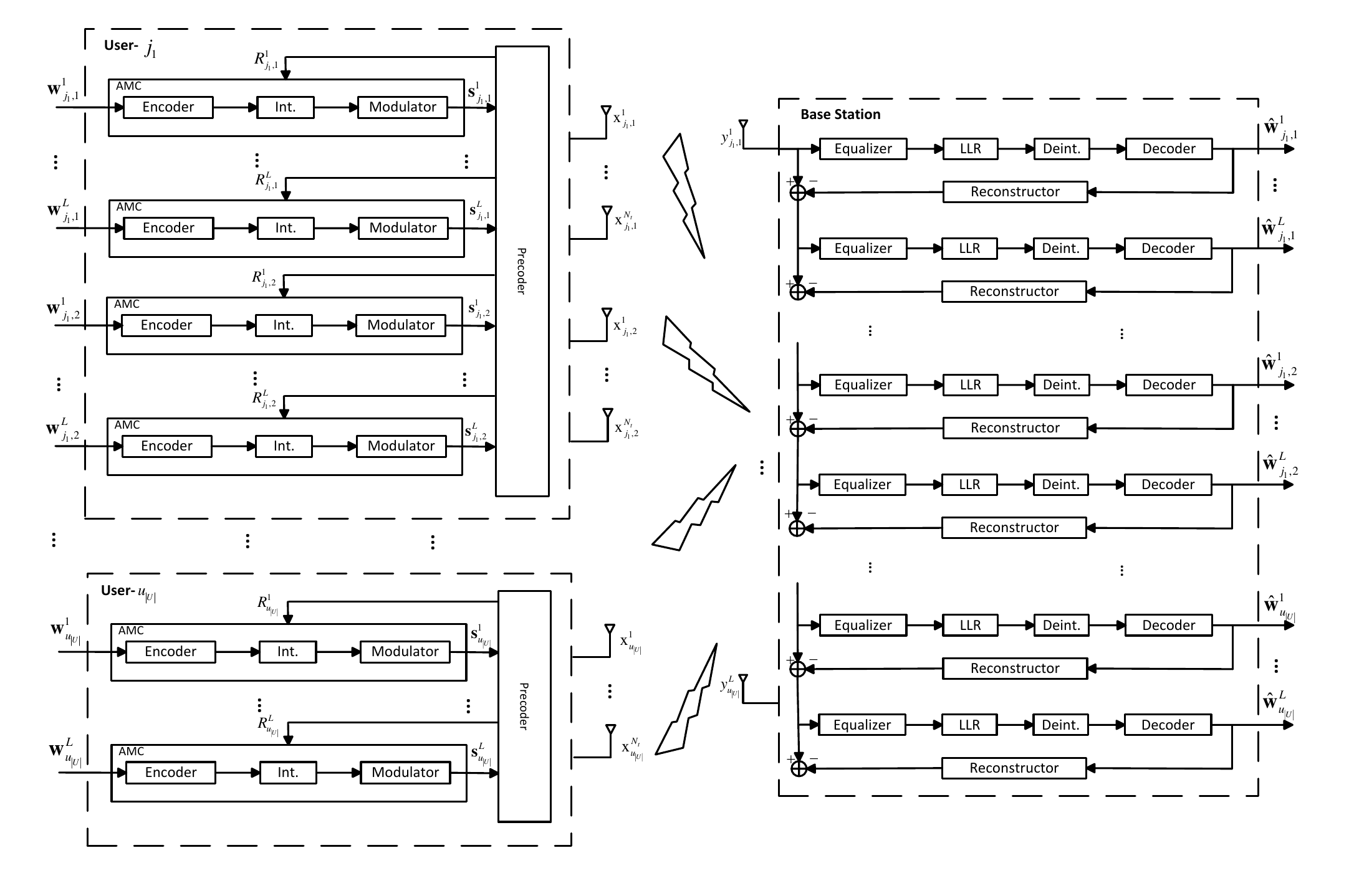}
    \caption{The Proposed Transceiver Architecture of uplink MIMO RSMA for Link-Level Simulations.}
    \label{Fig.3}
\end{figure*} 
Apart from the theoretical foundations, it is crucial to prove the enhanced efficiency of RSMA in practical scenarios. Practical transceiver architectures for RSMA in downlink MISO and MIMO channels have been designed in \cite{dizdar2020rate} and \cite{mishra2021rate}. In this section, we propose a practical transceiver architecture for uplink MIMO RSMA setting. Fig. \ref{Fig.3} illustrates the proposed transceiver architecture.

At the transmitter side, finite alphabet modulation schemes 4-QAM, 16-QAM, 64-QAM and 256-QAM, finite-length polar coding\cite{arikan2009channel} for Forward Error Correction (FEC) and an AMC algorithm are employed. The two messages from the splitting user-$j_{l}$ are mapped to bit vectors $\textbf{w}_{j_{l},d}^{a}$ of length ${K}_{j_{l},d}^{a}$, for $j_{l}\in\mathcal{J}, ~d\in\{1,2\}$ and $a\in\{1, ..., L\}$. Similarly, the messages from the non-splitting user-$u_{q}$ are mapped to bit vectors $\textbf{w}_{u_{q}}^{b}$ of length ${K}_{u_{q}}^{b}$, for $u_{q}\in\mathcal{U}$ and $b\in\{1, ..., L\}$. Then, the information bit vectors of splitting user-$j_{l}$ and non-splitting user-$u_{q}$ are independent and uniformly distributed in $\mathbb{F}_{2}^{{K}_{j_{l},d}^{a}}$ and $\mathbb{F}_{2}^{{K}_{u_{q}}^{b}}$, respectively. 
Then the information bit vectors are independently encoded, interleaved and modulated into symbol streams $\textbf{s}_{j_{l},d}^{a}$ and $\textbf{s}_{u_{q}}^{b}$, each of length $S$. The modulation and coding parameters are determined by an AMC algorithm according to the corresponding rate. The precoders for all streams are obtained as described in Algorithm 1. More details on the channel coding procedure and AMC algorithm are given in \cite{dizdar2020rate} for interested readers.

At the receiver side, instead of performing joint decoding which has high complexity in implementations, interference nulling and interference cancellation are utilized to benefit from low-complexity decoding methods. The received signal is first detected using an equalizer. Minimum Mean Square Error (MMSE) equalizer is applied here. Log-Likelihood Ratios (LLR) are calculated from the combined signal for Soft Decision (SD) decoding of polar codes. Next, the LLRs are de-interleaved and decoded to acquire the decoded bits. The reconstruction process involves repeating operations at the transmitter for the decoded bits to obtain a precoded signal and multiplying this signal with the corresponding channel matrix.

According to the decoding order proposed in Sec. \ref{2.1}, when decoding the first symbol vector $\textbf{s}_{{j}_{1},1}$, containing $L$ streams, we assume the first $a-1$, $a\leq L$ streams have been correctly decoded and subtracted from the received signal. Then the remaining signal $\Tilde{\textbf{y}}^{\textbf{s}_{{j}_{{1},1}}^{a}}$ can be written as
\begin{equation} \label{eq.26}
\begin{split}
    \Tilde{\textbf{y}}^{\textbf{s}_{{j}_{1},1}^{a}} & = \textbf{H}_{j_{1}}\sum_{i=a}^{L}\textbf{p}_{j_{1},1}^{i}\textbf{s}_{j_{1},1}^{i}+\textbf{H}_{j_{1}}\textbf{P}_{j_{1},2}\textbf{s}_{j_{1},2} \\
    & \sum_{j_{l,l\neq1}\in\mathcal{J}}\textbf{H}_{j_{l}}\sum_{d=1}^{2}\textbf{P}_{j_{l},d}\textbf{s}_{j_{l},d}+\sum_{u_{q}\in\mathcal{U}}\textbf{H}_{u_{q}}\textbf{P}_{u_{q}}\textbf{s}_{u_{q}}+\textbf{n},
\end{split}    
\end{equation}
where $\textbf{p}_{j_{1},1}^{i}\in\mathbb{C}^{N_{t}\times1}$ is the precoder of $i^{\text{th}}$ stream of $\textbf{s}_{j_{1},1}$, corresponding the $i^{\text{th}}$ column of precoder $\textbf{P}_{j_{1},1}$ and $\textbf{s}_{j_{1},1}^{i}$ represents the stream $i$ of symbol vector $\textbf{s}_{j_{1},1}$. The corresponding MMSE combiner is given by
\begin{equation} \label{eq.27}
    \textbf{g}^{\textbf{s}_{{j}_{1},1}^{a}}=\left(\textbf{H}_{j_{1}}\textbf{p}_{j_{1},1}^{a}\right)^{H}\left(\sum_{i=a}^{L}\textbf{H}_{j_{1}}\textbf{p}_{j_{1},1}^{i}{\textbf{p}_{j_{1},1}^{i}}^{H}\textbf{H}_{j_{1}}^{H}+\textbf{N}_{j_{1}}\right)^{-1},
\end{equation}
where 
\begin{equation} \label{eq.28}
    \begin{split}                                   
        \textbf{N}_{j_{1}} & =\textbf{H}_{j_{1}}\textbf{P}_{j_{1},2}\textbf{P}_{j_{1},2}^{H}\textbf{H}_{j_{1}}^{H}+\sum_{j_{l,l\neq1}\in\mathcal{J}} \sum_{d=1}^{2}\textbf{H}_{j_{l}}\textbf{P}_{j_{l},d}\textbf{P}_{j_{l},d}^{H}\textbf{H}_{j_{l}}^{H} \\
        & +\sum_{u_{q}\in\mathcal{U}} \textbf{H}_{u_{q}}\textbf{P}_{u_{q}}\textbf{P}_{u_{q}}^{H}\textbf{H}_{u_{q}}^{H}+I.
    \end{split}
\end{equation}
After combining the signal $\textbf{g}^{\textbf{s}_{{j}_{1},1}^{a}}\Tilde{\textbf{y}}^{\textbf{s}_{{j}_{1},1}^{a}}$, LLRs are calculated. We use the LLRs calculation method in \cite{seethaler2004efficient}. Let $\lambda_{i}^{\textbf{s}_{{j}_{1}}^{a}}$ denote the $i^{\text{th}}$ bit of the combined stream $\textbf{s}_{{j}_{1},1}^{a}$. We write 
\begin{equation} \label{eq.29}
    \lambda_{i}^{\textbf{s}_{{j}_{1},1}^{a}}=\gamma^{\textbf{s}_{{j}_{1},1}^{a}}[\min_{a\in\theta_{1}^{i}}\psi(a)-\min_{a\in\theta_{0}^{i}}\psi(a)],
\end{equation}
where $\gamma^{\textbf{s}_{{j}_{1},1}^{a}}$ is calculated by \eqref{eq.4} with $m=1$, $\theta_{b}^{i}$ is the set of modulation symbols with the value $b, ~b\in\{0, 1\}$ at the $i^{\text{th}}$ bit location, $\psi(a)=|\frac{\textbf{g}^{\textbf{s}_{{j}_{1},1}^{a}}\Tilde{\textbf{y}}^{\textbf{s}_{{j}_{1},1}^{a}}}{\phi_{\textbf{s}_{{j}_{1},1}^{a}}-a}|^{2}$ and $\phi_{\textbf{s}_{{j}_{1},1}^{a}}=\frac{\gamma^{\textbf{s}_{{j}_{1},1}^{a}}}{1+\gamma^{\textbf{s}_{{j}_{1},1}^{a}}}$.
The similar procedure is applied to all the remaining streams. The decoding operation utilizes a Successive Cancellation List (SCL) decoder for point-to-point polar codes\cite{tal2015list}.

\section{Numerical Results and Disucssion} \label{5}
The performance of FBL RSMA with perfect CSIT and CSIR is evaluated and compared with conventional transmission schemes in this section. We illustrate the MMF performance of uplink MIMO RSMA and followed by the LLS results. Here, we compare the performance of RSMA to NOMA and SDMA.
\begin{itemize}  
\item NOMA: This is a special case of RSMA. As specified in Sec. \ref{2.2}, none of the users split their messages in NOMA. The BS sequentially decodes the user messages based on SIC. It is applied by expanding the precoder technique suggested in\cite{xu2024max} to the MIMO scenario with combiner optimization mentioned in Sec. \ref{3.2}.
\item SDMA: According to \cite{mao2018rate}, SDMA is a classical method where none of the users split their messages, and the BS decodes the message of each user by treating the messages of all other users as interference. 
\end{itemize}
We perform simulations for Rayleigh Fading channel with 100 channel realizations. The tolerance of the algorithm is set to be $\tau=10^{-4}$.
\subsection{MMF Performance}\label{4.1}
\begin{figure}[t]
    \centering
    \subfigure[$N_{r}=4$]{
        \includegraphics[scale=0.55]{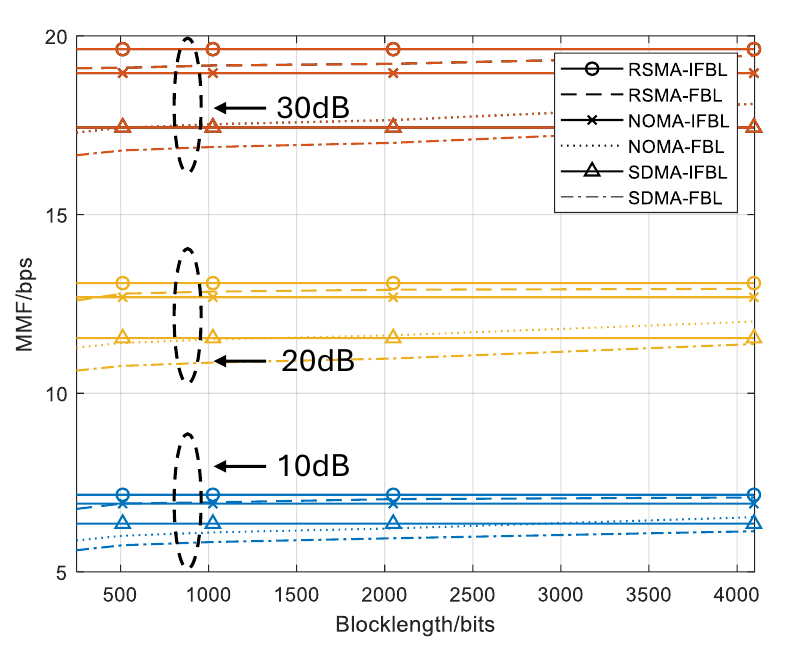}
        \label{Fig.4(a)}
    }
    \subfigure[$N_{r}=2$]{
	\includegraphics[scale=0.55]{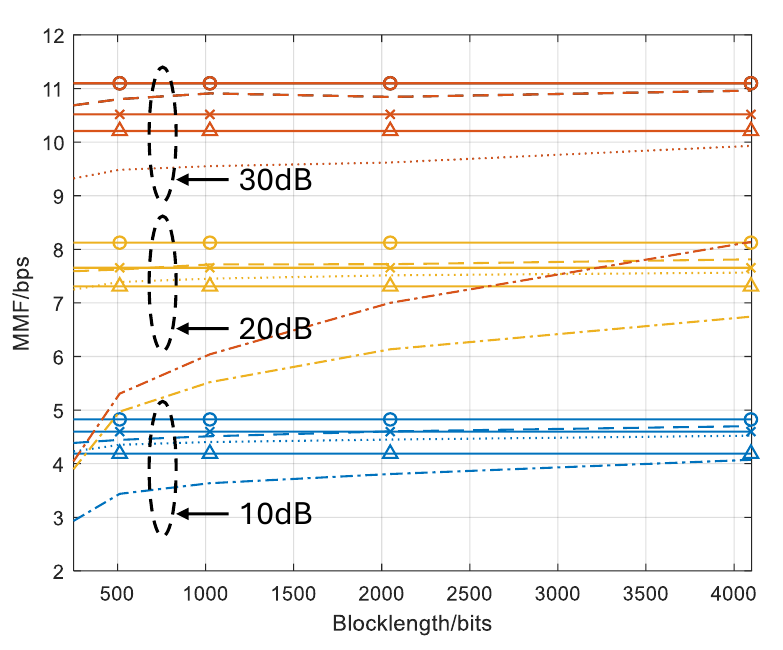}
        \label{Fig.4(b)}
    }
    \caption{The MMF performance of RSMA, NOMA and SDMA versus blocklengths with three different transmit SNR averaged over 100 random channel realizations. $K=2$. $N_{t}=2$. (a) is the underloaded scenario; (b) is the overloaded scenario.}
    \label{Fig.4}
\end{figure}

The trend of MMF as blocklength increases with $K=2$ is shown in Fig. \ref{Fig.4}. In this simulation, the number of transmit antenna $N_{t}$ is 2. Blue, yellow and red lines represent the transmit SNR of 10 dB, 20 dB and 30 dB, respectively. Fig. \ref{Fig.4(a)} shows the results of $N_r=4$, corresponding to the underloaded scenario while the overloaded scenario with $N_r=2$ is shown in Fig. \ref{Fig.4(b)}. 

It is evident that in both cases, FBL RSMA consistently outperforms FBL NOMA and FBL SDMA. With underloaded network settings in Fig. \ref{Fig.4(a)}, the benefits of FBL RSMA become significant when the transmit SNR is high, i.e., FBL RSMA with blocklength larger than 500 bits even outperforms IFBL NOMA with a transmit SNR of 30 dB.

In overloaded user settings in Fig. \ref{Fig.4(b)}, the performance gain of FBL RSMA over FBL SDMA becomes more significant as the transmit SNR increases due to its adeptness in managing overloaded scenarios. As evidence, given a transmit SNR of 30 dB, FBL SDMA with a 2000 bits blocklength (the red dashed curve with no marker) is even outperformed by FBL RSMA with a lower transmit SNR (20 dB) and a shorter FBL blocklength (e.g., 1500 bits) (the yellow dashed curve with circle marker). Besides comparison with FBL SDMA, the MMF rate for FBL RSMA with blocklengths of 1500 bits and 1000 bits surpasses that of IFBL NOMA when the transmit SNR is 10 dB and 20 dB, respectively. Since the decoding order strictly limits NOMA, the streams of the second decoded user would influence the SINR of the first decoded user significantly. However, with the assistance of RSMA, the decoding order can be more flexible to balance the SINR between each decoded stream.

\begin{figure}[t]
    \centering
    \subfigure[$N_{r}=8$]{
        \includegraphics[scale=0.65]{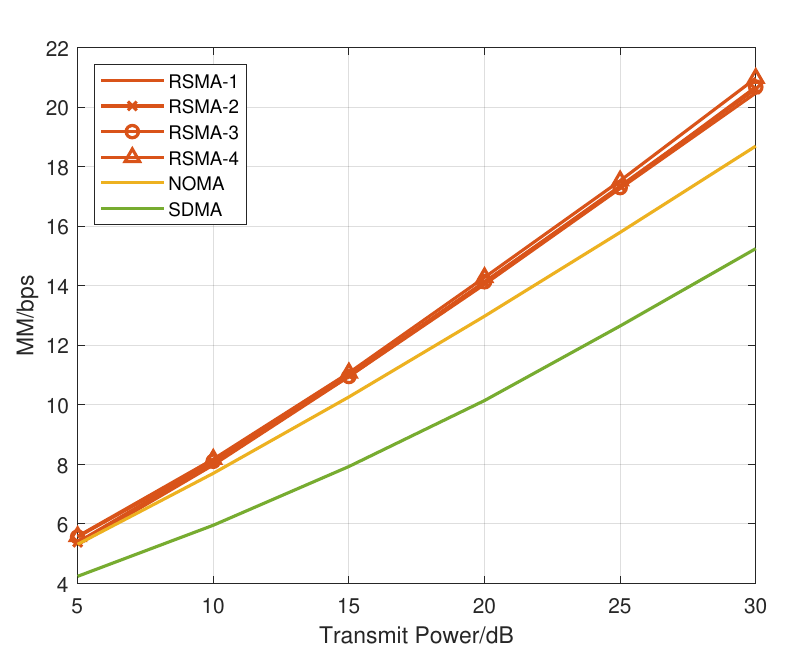}
        \label{Fig.5(a)}
    }   
    \subfigure[$N_{r}=8$]{
        \includegraphics[scale=0.5]{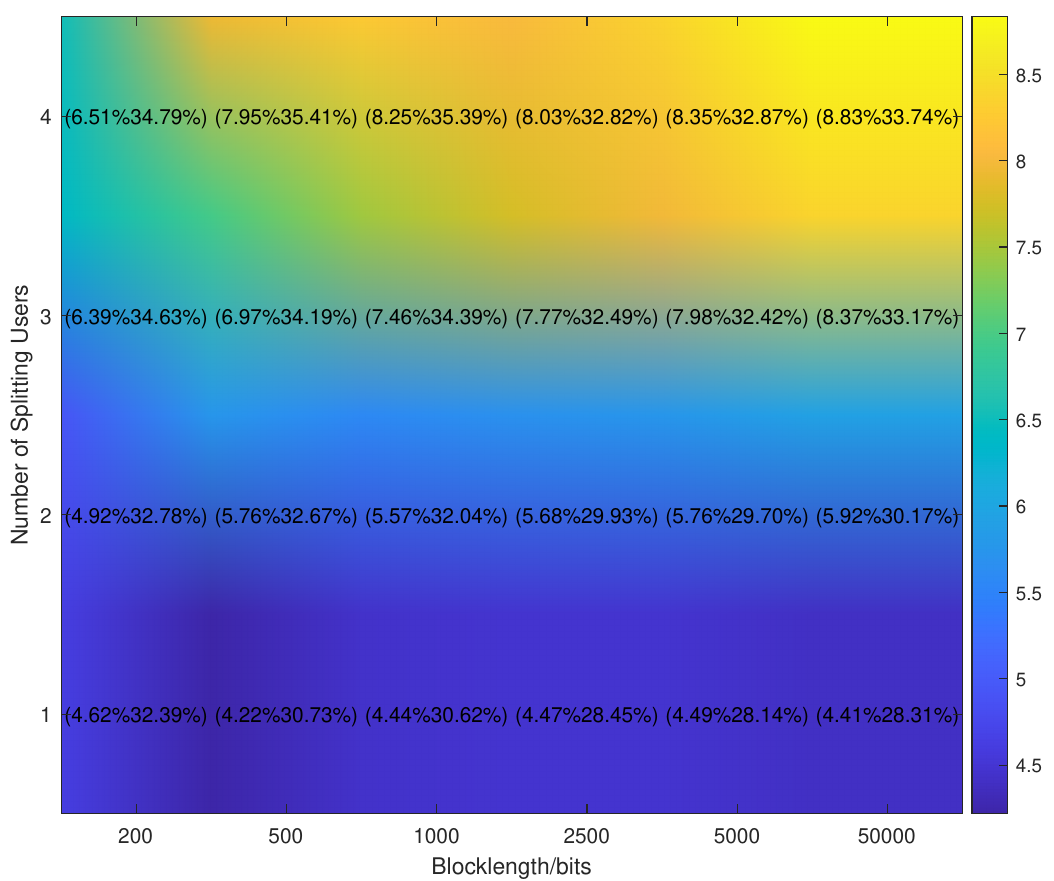}
        \label{Fig.5(b)}
    }
    \caption{The MMF performance of RSMA, NOMA and SDMA in the underloaded network settings averaged over 100 random channel realizations. Blocklength is 250 bits. $K=4$. $N_{t}=2$. (a) is the MMF versus transmit SNR; (b) is the relative gain of RSMA at the transmit SNR of 20 dB.}
    \label{Fig.5}
\end{figure}

\begin{figure}[t]
    \centering
    \subfigure[$N_{r}=4$]{
        \includegraphics[scale=0.65]{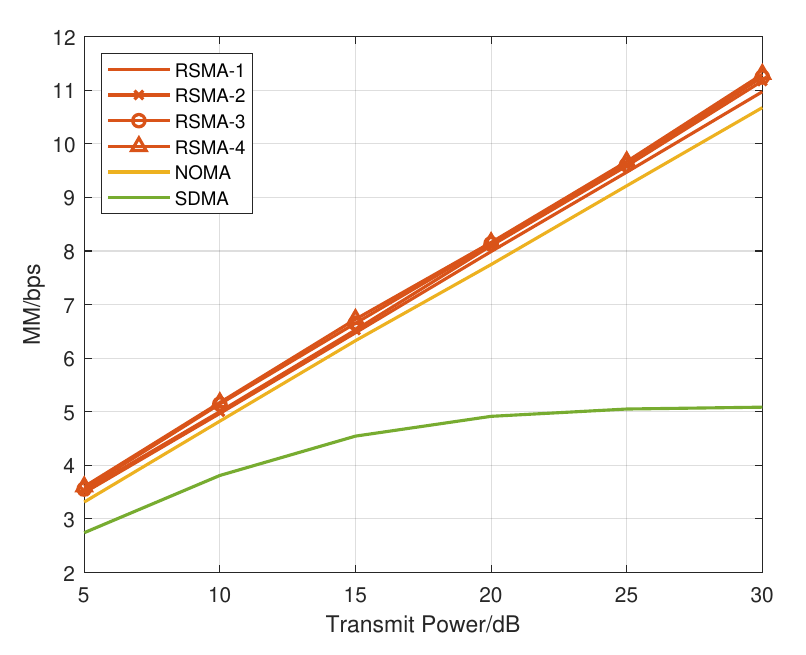}
        \label{Fig.6(a)}
    }   
    \subfigure[$N_{r}=4$]{
        \includegraphics[scale=0.5]{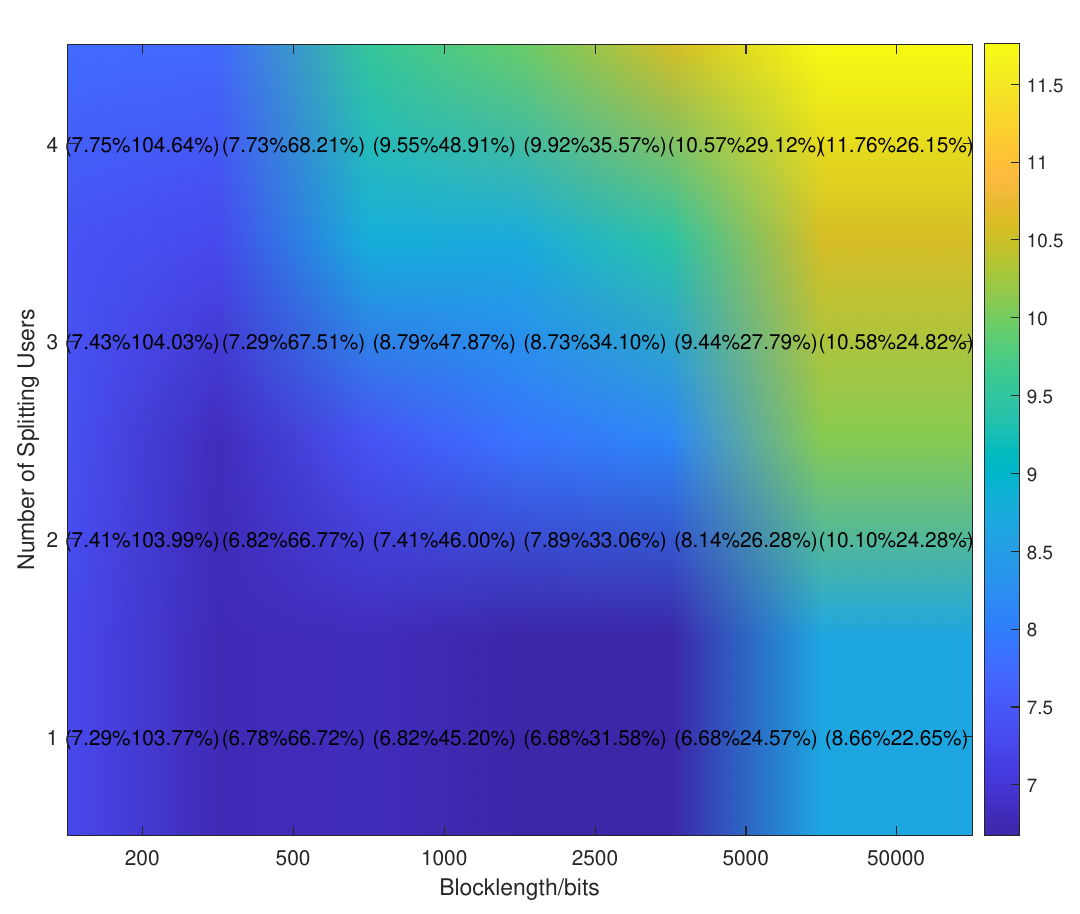}
        \label{Fig.6(b)}
    }
    \caption{The MMF performance of RSMA, NOMA and SDMA in the overloaded network settings averaged over 100 random channel realizations. Blocklength is 250 bits. $K=4$. $N_{t}=2$. (a) is the MMF versus transmit SNR; (b) is the relative gain of RSMA at the transmit SNR of 20 dB.}
    \label{Fig.6}
\end{figure}

In Fig. \ref{Fig.4}, we only take into account splitting one user in the two-user scenario as it has been proved that splitting both users has the same effect as splitting one user \cite{liu2020rate}. We now examine the effect of the number of splitting users when $K=4$. We compare the MMF of NOMA and SDMA to: \romannumeral1) RSMA with one splitting user; \romannumeral2) RSMA with two splitting users; \romannumeral3) RSMA with three splitting users, and \romannumeral4) RSMA with four splitting users in both underloaded and overloaded cases with FBL constraints in Fig. \ref{Fig.5(a)} and Fig. \ref{Fig.6(a)}, respectively. The blocklength is $250$ bits. The relative gain of FBL RSMA compared to FBL NOMA and FBL SDMA is defined by \eqref{eq.30} as 
\begin{equation} \label{eq.30}
    \frac{R^{\text{RSMA}}_{\text{MMF}}-\{R^{\text{NOMA}}_{\text{MMF}},R^{\text{SDMA}}_{\text{MMF}}\}}{\{R^{\text{NOMA}}_{\text{MMF}},R^{\text{SDMA}}_{\text{MMF}}\}}\times100\%.
\end{equation}
Fig. \ref{Fig.5(b)} and Fig. \ref{Fig.6(b)} illustrate the gain of FBL RSMA over FBL NOMA at transmit SNR of 20 dB. The percentages in parentheses represent MMF rate gains of FBL RSMA over FBL NOMA and FBL SDMA (with the same blocklength).

In Fig. \ref{Fig.5(a)}, with underloaded network settings, FBL RSMA has the best performance among the three transmission schemes. The MMF of FBL RSMA with only one splitting user already has a gain of approximately 4.5\% and 30\%  over that of FBL NOMA and FBL SDMA, respectively shown in Fig. \ref{Fig.5(b)}. Besides, we note that FBL RSMA with more splitting users results in greater MMF. Specifically, increasing the number of splitting users from 1 to 4 brings a fairness gain of around 7\% and 34\% over FBL NOMA and SDMA.

The overloaded case is illustrated in Fig. \ref{Fig.6}. Similarly to the case with $K=2$, the performance of FBL SDMA deteriorates greatly in Fig. \ref{Fig.6(a)}. As shown in Fig. \ref{Fig.6(b)}, FBL RSMA has around 100\% gain over FBL SDMA when the blocklength is short (e.g. 200 bits). Compared to the underloaded scenario, the gain of FBL RSMA over FBL NOMA and FBL SDMA increases in the overloaded scenario no matter with short or long blocklength.

\subsection{Link-Level Simulation Results}
The LLS is performed in this section to analyze the max-min throughput performance of FBL RSMA and compare it with that of FBL NOMA and FBL SDMA, employing the practical transceiver architecture described in Sec. \ref{4}. We consider an uplink MIMO with 2 users and each has $N_{t}=2$ transmit antennas. The number of message stream is $L = \min\{N_t, N_r\}$.

\begin{figure}[t]
    \centering
    \subfigure[$N_r=4$]{
        \includegraphics[scale=0.55]{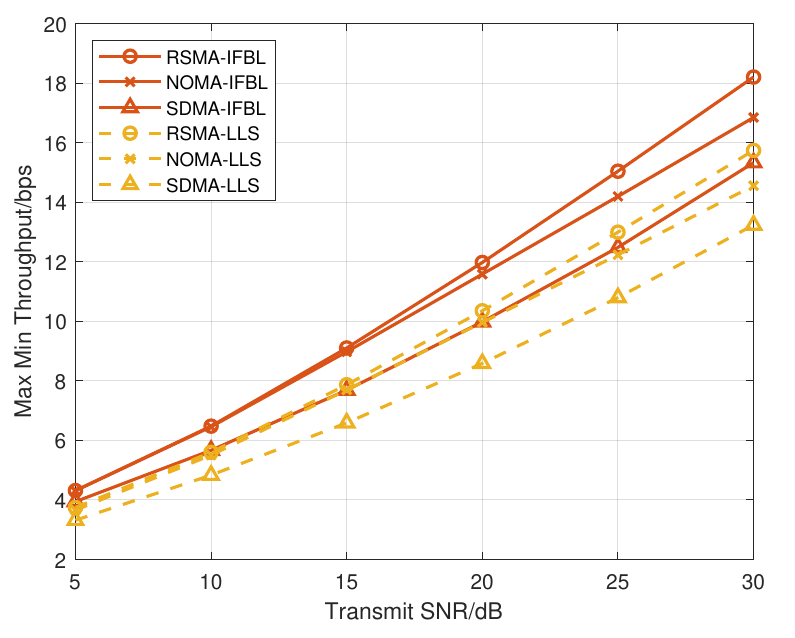}
        \label{Fig.7(a)}
    }   
    \subfigure[$N_r=2$]{
        \includegraphics[scale=0.55]{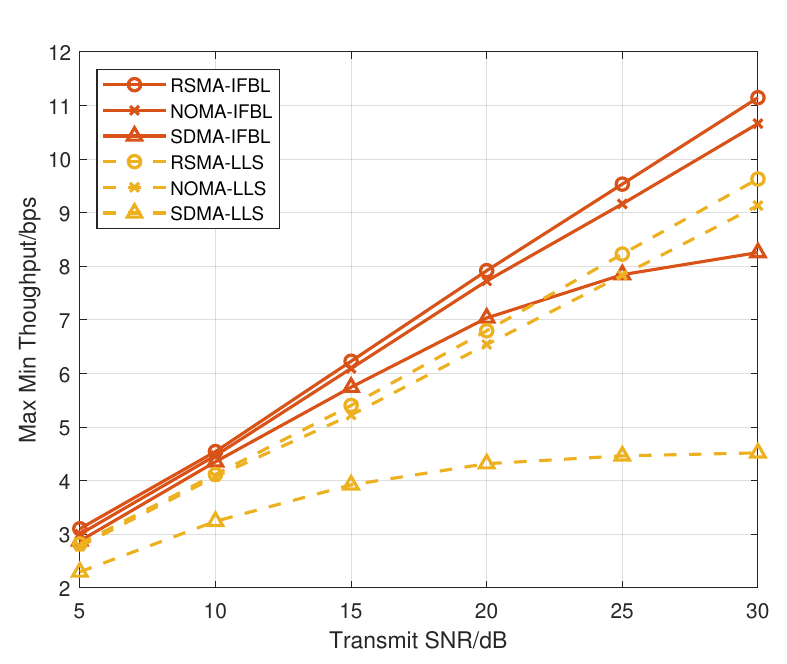}
        \label{Fig.7(b)}
    }
    \caption{The max-min throughput of RSMA, NOMA and SDMA versus transmit SNR averaged over 100 random channel realizations. $K=2$. (a) is the underloaded scenario; (b) is the overloaded scenario.}
    \label{Fig.7}
\end{figure}

With the assumption of perfect CSIT and CSIR for Modulation and Coding Scheme (MCS) selection, we use $S^{(l)}$, $D_{k}^{(l)}$ to denote the number of channel uses in the $l^{\text{th}}$ Monte-Carlo realization and the number of successfully recovered information bits by user-$k$. Then, we calculate the throughput as 
\begin{equation}\label{eq.31}
    \text{Max-Min Throughput[bps/Hz]}=\frac{\sum_{l}\min_{k\in\mathcal{K}}D_{k}^{(l)}}{\sum_{l}S^{(l)}}.
\end{equation}
We fix the modulated blocklength $S^{l}=256$ bits throughout the simulations. Fig. \ref{Fig.7} shows the MMF with IFBL and max-min throughput levels achieved by FBL RSMA, FBL NOMA and FBL SDMA in both overloaded and underloaded scenarios for $N_{t}=2$ and $K=2$. The max-min throughput performance in LLS is consistent with the MMF performance for all three schemes in both underloaded and overloaded regimes that FBL RSMA achieves a significant max-min throughput gain over FBL NOMA and FBL SDMA. However, with the consideration of coding and modulation, the theoretical max-min rate is not achievable, i.e., the throughput of FBL RSMA cannot outperform that of IFBL NOMA as it is the case in Sec. \ref{4.1}.

\section{Conclusion} \label{6}
To conclude, a general $K$-user RSMA framework in uplink MIMO with perfect CSIT and CSIR is introduced in this paper, where multiple streams are transmitted by users compared to only one stream transmitted in the SISO case. We investigate the benefits brought by RSMA with FBL constraints compared to NOMA and SDMA in terms of max-min user rate. Specifically, we propose an AO-based method to alternatively optimize the precoders at the transmitter side and the combiners at the receiver side, with each solved by SCA. We also apply a low-complexity decoding order method at the receiver side. Extensive numerical results demonstrate that FBL RSMA can obtain a greater MMF than FBL NOMA and FBL SDMA with the same blocklength. With the increase of blocklength, FBL RSMA can perform better than IFBL NOMA and IFBL SDMA. In multiple-user systems, FBL RSMA with one splitting user can have relatively high gain over FBL NOMA and FBL SDMA. The gain of RSMA becomes significant under the overloaded network settings. With multiple users being split, the gain of FBL RSMA increases. Moreover, the PHY-layer architecture of RSMA in uplink is designed and the performance is analyzed in practice. According to the LLS results, we demonstrate that FBL RSMA in practical settings can still outperform FBL NOMA and SDMA in MIMO. 

In summary, RSMA consistently achieves better user fairness than NOMA regardless of FBL or IFBL and RSMA can provide great performance. Increasing the number of splitting users can improve the performance of RSMA at the cost of more sophisticated hardware and processing. In addition, RSMA can achieve the same MMF as NOMA and SDMA but at a shorter blocklength, which achieves significantly lower latency. Therefore, we conclude that FBL RSMA is a strong and promising physical-layer approach for networks with multiple antennas.

\bibliographystyle{ieeetr}

\bibliography{ref.bib}

\end{document}